\documentclass{article}

\RequirePackage{latexsym,amssymb,amsfonts}
\RequirePackage{epsf}

\def\abstract#1{\vskip 7mm 
        \begin{center}{\large Abstract}\par \smallskip
                \begin{minipage}[c]{12cm}
                        \small #1
                \end{minipage}
        \end{center}
}
\def\title#1{\begin{center}{\Large\bf #1}\end{center}}
\def\author#1{\vskip 5mm \begin{center}{#1}\end{center}}
\def\address#1{\begin{center}{\it #1}\end{center}}
\makeatletter
\@ifundefined{lesssim}{}{}
\@ifundefined{gtrsim}{}{}
\def\vereq#1#2{\lower3pt\vbox{\baselineskip1.5pt \lineskip1.5pt
\ialign{$\m@th#1\hfill##\hfil$\crcr#2\crcr\sim\crcr}}}
\makeatother

\begin{document}

\title{Post-Newtonian Theory and its Application\footnote{To appear in
the Proc. of the 12th Workshop on General Relativity and Gravitation,
M. Shibata (ed.)}}

\author{Luc BLANCHET\footnote{E-mail address: blanchet@iap.fr}}
\address{Gravitation et Cosmologie (${\cal GR}\varepsilon{\cal CO}$),
\\ Institut d'Astrophysique de Paris, \\ 98$^{\rm bis}$ boulevard
Arago, 75014 Paris, France }

\abstract{We review some recent works on the post-Newtonian theory of
slowly-moving (post-Newtonian) sources, and its application to the
problems of dynamics and gravitational radiation from compact binary
systems. Our current knowledge is 3PN on the center-of-mass energy and
3.5PN on the gravitational-wave flux of inspiralling compact
binaries. We compute the innermost circular orbit (ICO) of binary
black-hole systems and find a very good agreement with the result of
numerical relativity. We argue that the gravitational dynamics of two
bodies of comparable masses in general relativity does not resemble
that of a test particle on a Schwarzschild background. This leads us
to question the validity of some ``Schwarzschild-like'' templates for
binary inspiral which are constructed from post-Newtonian resummation
techniques.}

\section{Introduction}\label{sec1}

Recent years have shown a tremendous revival of interest in an Old
Lady: the post-Newtonian approximation (or expansion when the speed of
light $c\to +\infty$), which is surely the most important technique in
the arsenal of General Relativity for drawing firm predictions for the
outcome of experimental facts related to gravitation. The
post-Newtonian approximation is ideally suited for describing the
adiabatic phase of the very interesting astrophysical systems known as
inspiralling compact binaries. These systems constitute our only
known-to-exist (for sure) source to hunt for in the current network of
laser-interferometric detectors of gravitational waves, today composed
of the large-scale interferometers VIRGO and LIGO, and the
medium-scale ones GEO and TAMA.

Two compact (i.e. gravitationally-condensed) objects --- neutron stars
or black holes --- orbit an inward spiral, with decreasing orbital
radius $r$, decreasing orbital period $P$, and increasing orbital
frequency $\omega=\frac{2\pi}{P}$. The inspiral is driven by the loss
of energy associated with the gravitational-wave emission, or,
equivalently, by the action of radiation forces. There is a long phase
of {\it adiabatic} inspiral, with associated dimensionless adiabatic
parameter

\begin{equation}\label{1}
\frac{\dot\omega}{\omega^2}={\cal O}\left(\frac{1}{c^5}\right)\;.
\end{equation}
The order of magnitude of the right-hand-side is that of the
radiation-reaction force, namely $\sim 1/c^5$ or equivalently 2.5PN
order beyond the standard Newtonian acceleration (following the usual
post-Newtonian jargon). The binary's dynamics being essentially
``aspherical'', inspiralling compact binaries are strong emitters of
gravitational radiation.

The main point about the theoretical description of inspiralling
compact binary is that a model made of two structureless
point-particles, characterized solely by two mass parameters $m_1$ and
$m_2$ (and possibly two spins), is sufficient. Most of the
non-gravitational effects usually plaguing the dynamics of binary star
systems: a magnetic field, an interstellar medium, etc., are dominated
by gravitational forces. However, the real justification for a model
of point particles is that the effects due to the finite size of the
compact bodies are small. In particular, the tidal interactions
between the two compact objects are expected to play a little role
during most of the inspiral phase; the mass transfer (in the case of
neutron stars) does not occur until very late, near the final
coalescence. Thus the inspiralling compact binaries are very clean
systems, essentially dominated by gravitational forces during most of
their life. This is the reason why these systems are so interesting
for relativity theorists!

Our fascination for the old post-Newtonian lady is that it was
recognized that {\it improved} waveform modelling is crucial for
constructing efficient templates for searching and measuring the
gravitational waves from inspiralling compact binaries in the
LIGO/.../TAMA network. By {\it improved} we mean in fact {\it much}
improved with respect to what is known from the ``Newtonian''
approximation, which corresponds in the case of the radiation field to
the Einstein quadrupole formalism. And one needs much improved
modelling because the orbital motion of the inspiralling compact
binary --- which is responsible for the gravitational-wave emission
--- becomes {\it very} relativistic when the binary reaches the
so-called Innermost Circular Orbit or ICO (defined below in this
article). At that point the orbital velocity is of the order of 50\%
of the speed of light. The price to be paid for applying the
post-Newtonian approximation is that we must go to very {\it high}
post-Newtonian order. The status of the field nowadays is the 3PN
approximation (i.e. $\sim 1/c^6$) --- or even better the 3.5PN one
---, which is likely to be sufficient for practical purpose (at least
in the case of neutron stars binaries).

After the two objects have passed the ICO they will plunge together
and merge to form a single black hole, which will subsequently settle
down into a stationary configuration, by emission of gravitational
waves in the quasi-normal mode channel. Of course we should not expect
the post-Newtonian approximation (and the description of the compact
objects by point-particles) to be valid after the ICO, and we shall
have to replace this model by a fully relativistic numerical
computation of the phase of plunge and merger of two black holes
\cite{numerical}. But, up to the point of the ICO, the post-Newtonian
approximation, when carried out to 3PN order, {\it is} physically
valid, and is probably very accurate. This last point may appear to be
a little surprising (it contradicts some statements in the litterature
about the ``failure'' of the post-Newtonian approximation in the
regime of the ICO), but we shall present arguments supporting it. In
particular the 3PN order is probably able to locate the ICO to within
an accuracy of the order of 1\% or better (for binary systems with
comparable masses).

\section{Post-Newtonian iteration for isolated systems}\label{sec2}

Before embarking into strong statements concerning the ``accuracy'' of
the post-Newtonian description of binary systems, it is wise to
examine in a somewhat general way the well-definiteness of the
approximation itself. This is especially important in view of the fact
that our old lady the post-Newtonian approximation has been unjustly
accused of being plagued with some apparently inherent difficulties,
and furthermore which crop up around the 3PN order we are interested
in. The problems arise in the general case of regular
(singularity-free) matter sources. Up to the 2.5PN order the
approximation can be worked out without problems, and often at the 3PN
order the problems can be solved specifically for each case at
hands. However, it must be admitted that these difficulties, even if
appearing at higher approximations, have cast doubt in the past on the
actual soundness, on the theoretical point of view, of the
post-Newtonian expansion. They pose the practical question of the
reliability of the approximation when comparing the theory's
predictions with very precise experimental results. Here we discuss
the nature of the problems -- are they purely technical or linked with
some fundamental drawback of the approximation? -- and outline their
resolution recently proposed in Ref. \cite{PB02}.

The first problem is that in higher approximations some {\it
divergent} Poisson-type integrals appear. Recall that the
post-Newtonian expansion replaces the resolution of an hyperbolic-like
d'Alembertian equation by a perturbatively equivalent hierarchy of
elliptic-like Poisson equations. Rapidly it is found during the
post-Newtonian iteration that the right-hand-side of the Poisson
equations acquires a non-compact support (it is distributed over all
space), and that the standard Poisson integral diverges because of the
bound of the integral at spatial infinity, i.e. $r\equiv |{\mathbf
x}|\to +\infty$, with $t=$ const. For instance some of the potentials
occuring at the 2PN order in Chandrasekhar's work \cite{CN69} are
divergent, so the corresponding metric is formally infinite. In fact,
Kerlick \cite{Ker80,Ker80'} showed that the post-Newtonian computation
{\it \`a la} Chandrasekhar, following the iteration scheme of Anderson
and DeCanio \cite{AD75}, can be made well-defined up to the 2.5PN
order, but that this does not solve the problem at the next 3PN order,
which has been found to involve some inexorably divergent Poisson
integrals.

These divergencies come from the fact that the post-Newtonian
expansion is actually a singular perturbation, in the sense that the
coefficients of the successive powers of $1/c$ are not uniformly valid
in space, since they typically blow up at spatial infinity like some
positive powers of $r$. For instance, Rendall \cite{Rend92} has shown
that the post-Newtonian expansion cannot be ``asymptotically flat''
starting at the 2PN or 3PN level, depending on the adopted coordinate
system. The result is that the Poisson integrals are in general
badly-behaving at infinity. Trying to solve the post-Newtonian
equations by means of the standard Poisson integral does not {\it a
priori} make sense. This does not mean that there are no solution to
the problem, but simply that the Poisson integral does not constitute
the correct solution of the Poisson equation in the context of
post-Newtonian expansions. So the difficulty is purely of a technical
nature, and will be solved once we succeed in finding the appropriate
solution to the Poisson equation\footnote{The problem is somewhat
similar to what happens in Newtonian cosmology. Here we have to solve
the Poisson equation $\Delta U=-4\pi G\rho$, where the density $\rho$
of the cosmological fluid is constant all over space:
$\rho=\rho(t)$. Clearly the Poisson integral of a constant density
does not make sense, as it diverges at the bound at infinity like the
integral $\int r dr$. This nonsensical result has occasionally been
referred to as the ``paradox of Seeliger''.  However the problem is
solved once we realize that the Poisson integral does not constitute
the appropriate solution of the Poisson equation in the context of
Newtonian cosmology. A well-defined solution is simply given by
$U=-\frac{2}{3}\pi G\rho r^2$.}.

To cure the problem of divergencies we have introduced \cite{PB02}, at
any post-Newtonian order, a generalized solution of the Poisson
equation with non-compact support source, in the form of an
appropriate {\it finite part} of the usual Poisson integral: namely we
regularize the bound at infinity of the Poisson integral by means of a
process of analytic continuation, analogous to the one already used to
regularize the retarded integrals in
Refs. \cite{BD86,B93,B95,B98mult}. At any post-Newtonian order $n$ we
have to solve a Poisson equation with non-compact-support
``source-term'' $\overline{\tau}^{\mu\nu}_n$, where the overbar
indicates that we are considering a (formal) post-Newtonian expansion,
and pick up the coefficient of $1/c^n$ in that expansion ($\mu,\nu$
are space-time indices). We multiply the source-term by a
regularization factor $r^B$, where $r=|{\mathbf x}|$ and $B\in{\mathbb
C}$, and then apply the standard Poisson integral. The result,

\begin{equation}\label{2}
\Delta^{-1}\left[r^B\overline{\tau}^{\mu\nu}_n\right]({\mathbf x},t) =
-\frac{1}{4\pi} \int_{{\mathbb R}^3} ~\frac{d^3{\mathbf y}}{|{\mathbf
x}-{\mathbf y}|} ~\!|{\mathbf
y}|^B~\!\overline{\tau}^{\mu\nu}_n({\mathbf y},t) \;, 
\end{equation}
is a well-defined integral (i.e. convergent at infinity, $r\to
+\infty$) in some region of the $B$-complex plane, given by $\Re
(B)<-a_\mathrm{max}-2$, where $a_\mathrm{max}$ denotes the maximal
power of $r$ in the behaviour of the source-term
$\overline{\tau}^{\mu\nu}_n$ when $r\to +\infty$, i.e. it corresponds
to the maximal growth of the source-term at infinity (notice that
$a_\mathrm{max}$ gets larger and larger when we increase the
post-Newtonian order $n$). Next, we can prove that the latter function
of $B$ generates a (unique) analytic continuation down to a
neighbourhood of the origin $B=0$, except at $B=0$ itself, around
which value it admits a Laurent expansion with multiple poles up to
some finite order. Then, we consider the Laurent expansion of that
function when $B\to 0$ and compute the finite part (${\cal FP}$), or
coefficient of the zero-th power of $B$, of that expansion. This
defines our generalized Poisson integral:

\begin{equation}\label{3}
\widetilde{\Delta^{-1}}\big[~\!\overline{\tau}^{\mu\nu}_n\big] = {\cal
FP}\Delta^{-1}\left[r^B\overline{\tau}^{\mu\nu}_n\right]
\;. \end{equation} The resulting ``generalized'' Poisson integral
constitutes an appropriate solution of the post-Newtonian
equation. However this is only a {\it particular} solution of the
Poisson equation, and the most general solution will be the sum of
that particular solution and the most general solution of the
corresponding homogeneous equation. At this stage, considering the
post-Newtonian iteration scheme alone, we cannot do more and therefore
we leave the homogeneous solution unspecified (we can see {\it a
posteriori} that it is associated with radiation-reaction effects).

The second problem has to do with the {\it near-zone} limitation of
the post-Newtonian approximation. Indeed the post-Newtonian expansion
assumes that all retardations $r/c$ are small, so it can be viewed as
a formal {\it near-zone} expansion when $r\to 0$, which is valid only
in the region surrounding the source that is of small extent with
respect to the typical wavelength of the emitted radiation: $r\ll
\lambda$ (if we locate the origin of the coordinates $r=0$ inside the
source). Therefore, the fact that the coefficients of the
post-Newtonian expansion blow up at spatial infinity, when $r\to
+\infty$, has nothing to do with the actual behaviour of the field at
infinity. The serious consequence is that it is not possible, {\it a
priori}, to implement within the post-Newtonian iteration the physical
information that the matter system is isolated from the rest of the
universe. Most importantly, the no-incoming radiation condition,
imposed at past null infinity, cannot be taken into account, {\it a
priori}, into the scheme.  In a sense the post-Newtonian approximation
is not ``self-supporting'', because it necessitates some information
taken from outside its own domain of validity.

The solution of the problem of the near-zone limitation of the
post-Newtonian expansion resides in the {\it matching} of the
near-zone field to the exterior field, a solution of the vacuum
equations outside the source which has been developed in previous
work \cite{BD86} using some post-{\it Minkowskian} and multipolar
expansions. In the case of post-Newtonian sources, the near zone,
i.e. $r\ll\lambda$, covers entirely the source, because the source's
radius itself is such that $a\ll\lambda$. Thus the near zone overlaps
with the exterior zone where the multipole expansion is valid. What we
do is to impose a matching condition resulting from the numerical
equality between the multipolar and post-Newtonian fields, verified in
the external part of the near-zone (say $a<r\ll\lambda$, where $a$ is
the size of the source). The matching equation reads

\begin{equation}\label{4}
\overline{{\cal M}(h^{\mu\nu})} \equiv {\cal M}(\overline{h}^{\mu\nu})\;,
\end{equation} 
where ${\cal M}$ symbolizes the multipolar (actually
multipolar-post-Minkowskian) series, and as before the overbar refers
to the post-Newtonian or near-zone expansion. We emphasize that the
matching equation is physically justified only for post-Newtonian
sources, for which the exterior near-zone exists.

The requirement of matching to the post-Newtonian solution has been
shown (in a previous work \cite{B98mult}) to entirely determine, up to
any post-Newtonian order, the multipole moments parametrizing the
exterior field. In our more recent paper \cite{PB02} we have proved
that the still undetermined homogeneous solutions in the
post-Newtonian iteration alluded to before are also fully determined
by the matching equation. These homogeneous solutions are associated
with radiation-reaction effects --- for instance they incorporate the
dominant radiation-reaction force at the 2.5PN order, as well as the
tail contribution in the radiation-reaction force which is known to
arise at the 4PN order \cite{BD88,B93,B97}). In conclusion, the
post-Newtonian expansion of the field inside an isolated system is
fully determined (it can be {\it indefinitely} reiterated, up to any
post-Newtonian order) by the matching to the exterior solution
satisfying the correct boundary condition at infinity --- notably the
absence of incoming radiation from past null infinity.

\section{Dynamics and radiation of compact binaries}\label{sec3}

By equations of motion we mean the explicit expression of the
accelerations of the compact bodies in terms of the positions and
velocities. In Newtonian gravity, writing the equations of motion for
a system of $N$ particles is trivial; in general relativity, even
writing the equations in the case $N=2$ is difficult. The first
relativistic term, at the 1PN order, was derived by Einstein, Infeld
and Hoffmann \cite{EIH} by means of their famous ``surface-integral''
method, by which the equations of motion are deduced from the {\it
vacuum} field equations, and so which is applicable to any compact
objects (they be neutron stars, black holes, or, maybe, naked
singularities).

Concerning the 2PN and 2.5PN approximations the result for the
equations of binary motion in harmonic coordinates was obtained by
Damour and Deruelle and collaborators
\cite{BeDD81,DD81a,DD81b,D82}. The corresponding result for the
ADM-Hamiltonian of two particles at the 2PN order was given in
Ref. \cite{DS85}. By ADM-Hamiltonian we refer to the Fokker-type
Hamiltonian, which is obtained from the matter-plus-field
Arnowitt-Deser-Misner Hamiltonian by eliminating the field degrees of
freedom. Kopeikin \cite{Kop85} derived the 2.5PN equations of motion
for two extended compact objects. The 2.5PN-accurate
harmonic-coordinate equations as well as the complete gravitational
field (namely the metric $g_{\mu\nu}$) were computed by Blanchet, Faye
and Ponsot \cite{BFP98}, and by Pati and Will \cite{PW02}.

It is important to realize that the 2.5PN equations of motion are
known to hold in the case of binary systems of strongly
self-gravitating bodies. This is {\it via} an ``effacing'' principle
for the internal structure of the bodies. As a result, the equations
depend only on the ``Schwarzschild'' masses, $m_1$ and $m_2$, of the
compact objects, as has been explicitly verified up to the 2.5PN order
\cite{Kop85,GKop86}. The 2.5PN equations of motion have also been
established by Itoh, Futamase and Asada \cite{IFA00,IFA01}, who use a
variant of the surface-integral approach \cite{EIH}, valid for compact
bodies.

The present state of the art is the 3PN order. The equations of motion
at this order have been worked out independently by two groups, by
means of different methods, and with equivalent results. On one hand,
Jaranowski and Sch\"afer \cite{JaraS98,JaraS99}, and Damour,
Jaranowski and Sch\"afer \cite{DJS00,DJS01,DJSdim}, employ the
ADM-Hamiltonian formalism of general relativity; on the other hand,
Blanchet and Faye \cite{BF00,BFeom,BFreg,BFregM}, de Andrade, Blanchet
and Faye \cite{ABF01}, and Blanchet and Iyer \cite{BI02CM}, founding
their approach on the post-Newtonian iteration initiated in
Ref. \cite{BFP98}, compute directly the equations of motion from which
they infer the Lagrangian (instead of the Hamiltonian) in harmonic
coordinates. The end results have been shown \cite{DJS01,ABF01} to be
physically equivalent in the sense that there exists a unique
``contact'' transformation of the dynamical variables, that changes
the harmonic-coordinates Lagrangian obtained in Ref. \cite{ABF01} into
a new Lagrangian, whose Legendre transform coincides exactly with the
Hamiltonian given in Ref. \cite{DJS00}. The 3PN equations of motion,
however, depend on one unspecified numerical coefficient, $\omega_{\rm
static}$ in the ADM-Hamiltonian formalism and $\lambda$ in the
harmonic-coordinates approach, which is due to some incompleteness of
the Hadamard self-field regularization method. This coefficient has
been fixed by means of a dimensional regularization in
Ref. \cite{DJSdim} (see Section \ref{sec5}).

So far the status of the post-Newtonian equations of motion is quite
satisfying. There is mutual agreement between all the results obtained
by means of different approaches and techniques, whenever it is
possible to compare them: point-particles described by Dirac
delta-functions, extended post-Newtonian fluids, surface-integrals
methods, mixed post-Minkowskian and post-Newtonian expansions, direct
post-Newtonian iteration and matching, harmonic coordinates versus
ADM-type coordinates, different processes or variants of the
regularization of the self field of point-particles. 

Let us remark that the 3PN equations of motion are merely
``Newtonian'' as regards the radiative aspects of the problem, because
with that precision they contain the radiation reaction force at only
the lowest 2.5PN order. From the conservative part of the 3PN dynamics
(neglecting the 2.5PN radiation reaction) we shall obtain the binary's
center-of-mass energy $E$ at the 3PN order. We now want to compute the
variation of $E$ because of the emission of gravitational
radiation. For this purpose we replace the knowledge of the radiation
reaction force in the local equations of motion by the computation of
the total energy flux in gravitational waves, say ${\cal L}$, and we
apply the energy balance equation

\begin{equation}\label{5}
\frac{d E}{dt}=-{\cal L}\;.
\end{equation}
Therefore in our approach the computation of the center-of-mass energy
$E$ constitutes only one ``half'' of the solution of the problem. The
second ``half', that of the computation of the energy flux ${\cal L}$,
is to be carried out by application of a wave-generation formalism.

Following earliest computations at the 1PN level \cite{WagW76,BS89}
(at a time when the post-Newtonian corrections in ${\cal L}$ had a
purely academic interest), the energy flux of inspiralling compact
binaries was completed to the 2PN order by Blanchet, Damour and Iyer
\cite{BDI95}, using the 2PN wave-generation formalism of
Ref. \cite{B95}, and, independently, by Will and Wiseman \cite{WWi96},
using their own formalism (see Refs. \cite{BDIWWi95,BIWWi96} for joint
reports of these calculations). The preceding approximation, 1.5PN,
which represents in fact the dominant contribution of tails in the
wave zone, had been obtained in Refs. \cite{Wi93,BS93} by application
of a formula for tail integrals given in
Ref. \cite{BD92}. Higher-order tail effects at the 2.5PN and 3.5PN
orders, as well as a crucial contribution of tails generated by the
tails themselves (the so-called ``tails of tails'') at the 3PN order,
were obtained by Blanchet \cite{B96,B98tail}. However, unlike the
1.5PN, 2.5PN and 3.5PN orders that are entirely composed of tail
terms, the 3PN approximation involves also, besides the tails of
tails, many non-tail contributions coming from the relativistic
corrections in the (source) multipole moments of the binary. These
have been completed by Blanchet, Iyer and Joguet
\cite{BIJ02,BFIJ02}, based on the expressions for the
multipole moments given in Ref. \cite{B98mult}, except for one single
unknown numerical coefficient, due to the use of the Hadamard
regularization, which is a combination of the parameter $\lambda$ in
the equations of motion, and a new parameter $\theta$ coming from the
computation of the 3PN quadrupole moment.

The post-Newtonian flux ${\cal L}$ is in complete agreement, up to the
3.5PN order, with the result given by the very different technique of
linear black-hole perturbations, valid in the ``test-mass'' limit
where the mass of one of the bodies tends to zero (limit $\nu\to 0$,
where $\nu=\mu/m$). Linear black-hole perturbations, triggered by the
geodesic motion of a small mass around the black hole, have been
applied to this problem \cite{P93,TNaka94}. This technique has
culminated with the beautiful analytical methods of Sasaki, Tagoshi
and Tanaka \cite{Sasa94,TSasa94,TTS96}, who solved the problem up to
the extremely high 5.5PN order.

\section{Post-Newtonian templates for compact binary inspiral}\label{sec4}

The orbital phase of the binary --- i.e. the integral of the frequency
$\omega(t)$: 

\begin{equation}\label{6}
\Phi(t) = \int \omega(t) dt\;,
\end{equation}
constitutes the crucial quantity to be monitored (and therefore to be
predicted) in the detectors. The templates built from the theoretical
prediction for the orbital phase should be accurate enough over most
of the inspiral phase, within the frequency bandwidth of the
detectors, with reduced cumulative phase lags, so that the phasing
errors are not significant when one attempts to extract the values of
the binary's parameters (essentially the masses and spins) from the
data.

The relevant model for describing the inspiral phase consists of two
point-masses moving under their mutual gravitational attraction. As a
simplification for the theoretical analysis, the orbit of inspiralling
binaries can be considered to be circular, apart from the gradual
inspiral, with a good approximation. The templates are based on the
energy-balance equation (\ref{5}), from which one deduces the orbital
phase as

\begin{equation}\label{7}
\Phi_c-\Phi=\int_{\omega_c}^\omega\frac{\omega dE}{{\cal L}}\;,
\end{equation}
where $\Phi_c$ and $\omega_c$ denote the values at the instant of
coalescence. The number of gravitational-wave cycles left from the
current time till the coalescence instant (we consider only the
dominant harmonics at twice the orbital frequency) is

\begin{equation}\label{8}
{\cal N}=\frac{\Phi_c-\Phi}{\pi}\;.
\end{equation}
It is clear --- because of Eq. (\ref{1}) --- that ${\cal N}$ is of the order
of the inverse of radiation-reaction effects, hence the formal
``post-Newtonian'' order is ${\cal N}={\cal O}\left(c^{+5}\right)$. 

As a matter of fact, ${\cal N}$ will be a large number, approximately
equal to $1.6\times 10^4$ in the case of two neutron stars between 10
and 1000~Hz (roughly the frequency bandwidth of the detector
VIRGO). Data analysts have estimated that, in order not to suffer a
too severe reduction of signal-to-noise ratio, one should monitor the
phase evolution with an accuracy comparable to one gravitational-wave
cycle (i.e. $\delta {\cal N}\sim 1$), over the whole detector's
bandwidth. From a strict post-Newtonian point of view, we see that the
consequence, since the ``Newtonian'' number of cycles is formally
${\cal O}\left(c^{+5}\right)$, is that any post-Newtonian correction
therein that is larger than the order $c^{-5}$ is expected to
contribute to the phase evolution more than what is allowed by the
previous estimate. Therefore, one expects that in order to construct
accurate enough templates it will be necessary to compute the orbital
phase up to at least the 2.5PN order at a minimum. This
back-on-the-envelope estimate has been confirmed by
measurement-accuracy analyses \cite{CFPS93,TNaka94,P95,DIS98} which
showed that in advanced generations of detectors the 3PN approximation
(or, even better, the 3.5PN one) is required in the case of
inspiralling neutron star binaries.

The first ingredient in the theoretical calculation is the binding
energy $E$ (in the center-of-mass frame), defined as being the
invariant energy associated with the {\it conservative} part of the
binary's 3PN dynamics (we ignore the radiation reaction effect at the
2.5PN order). The center-of-mass energy $E$ is deduced from the
Hamiltonian in ADM-type coordinates
\cite{JaraS98,JaraS99,DJS00,DJS01,DJSdim}, or equivalently from the
Lagrangian in harmonic-coordinates
\cite{BF00,BFreg,BFregM,BFeom,ABF01,BI02CM}. Restricting our
consideration to circular orbits, the energy is a function or a single
variable, the radial distance $r$ between the two particles in a given
coordinate system. In fact it is better to express the energy in terms
of the frequency $\omega=\frac{2\pi}{P}$ of the orbital motion, or,
rather, in terms of the particular frequency-related parameter

\begin{equation}\label{9}
x \equiv \left(\frac{G M \omega}{c^3}\right)^{2/3}\;.
\end{equation}
The individual masses of the black holes are denoted by $m_1$ and
$m_2$, and the total mass is $M=m_1+m_2$. The interest of using the
parameter $x$, instead of some coordinate distance $r$, is that the
energy function $E(x)$ then takes an invariant invariant (the same in
different coordinate systems). The result we get consists of the
Newtonian contribution, proportional to $x$, followed by
post-Newtonian corrections up to the 3PN order:

\begin{eqnarray}\label{10}
E &=& -\frac{\mu c^2 x}{2} \biggl\{ 1
 +\left(-\frac{3}{4}-\frac{1}{12}\nu\right) x +
 \left(-\frac{27}{8}+\frac{19}{8}\nu -\frac{1}{24}\nu^2\right) x^2
 \nonumber\\
 &+&\left(-\frac{675}{64}+\left[\frac{209323}{4032}-\frac{205}{96}\pi^2
 -\frac{110}{9}\lambda\right]\nu-\frac{155}{96}\nu^2
 -\frac{35}{5184}\nu^3\right)x^3 \biggr\}\;.
\end{eqnarray}
This expression involves the useful ratio between reduced and total
masses:

\begin{equation}\label{11}
\nu = \frac{\mu}{M} \quad\hbox{where}\quad \mu = \frac{m_1 m_2}{M}\;.
\end{equation}
This ratio is interesting because of its range of variation:
$0<\nu\leq \frac{1}{4}$, where $\nu=\frac{1}{4}$ in the equal-mass
case and $\nu\to 0$ in the test-mass limit for one of the bodies. The
parameter $\lambda$ denotes a point-mass regularization-constant and
will be discussed in Section \ref{sec5}.

The second ingredient in this analysis concerns the gravitational-wave
luminosity ${\cal L}$, calculated, in the post-Newtonian
approximation, from a wave-generation formalism valid for extended
``fluid'' systems \cite{B95,B98tail,B98mult}, and then specialized to
binary systems of point-particles
\cite{BDI95,B96,B98tail,BIJ02,BFIJ02}. The calculation takes properly
into account the relativistic corrections linked with the description
of the source (multipole moments), as well as the non-linear effects
in the propagation of the waves from the source to the far zone. The
crucial input of any post-Newtonian computation of the flux is the
mass quadrupole moment (because the post-Newtonian precision required
for the higher multipole moments is smaller). The 3PN quadrupole
moment for circular binary orbits in a harmonic coordinate system is
of the form\footnote{We neglect a 2.5PN term and denote e.g. ${\hat
x}_{ij}=x_ix_j-\frac{1}{3}\delta_{ij}r^2$, where $x_i$ is the
harmonic-coordinate separation between the two bodies and $r=|{\bf
x}|$.}

\begin{equation}\label{12}
I_{ij}=\mu\left(A~\!{\hat x}_{ij}+B~\!\frac{r^2}{c^2}~\!{\hat
v}_{ij}\right)\;,
\end{equation}
and has been obtained recently by Blanchet, Iyer and Joguet
\cite{BIJ02}, with result

\begin{eqnarray}
 A &=& 1+\left(-\frac{1}{42}-\frac{13}{14}\nu\right) \gamma +
 \left(-\frac{461}{1512}-\frac{18395}{1512}\nu
 -\frac{241}{1512}\nu^2\right) \gamma^2 \nonumber\\
 &+&\left(\frac{395899}{13200}-\frac{428}{105}\ln\left(\frac{r}{r_0}\right)
 +\left[\frac{139675}{33264}-\frac{44}{3}\ln\left(\frac{r}{r'_0}\right)
 -\frac{44}{3}\xi-\frac{88}{3}\kappa\right]\nu
 +\frac{162539}{16632}\nu^2+\frac{2351}{33264}\nu^3\right) \gamma^3\;,
 \nonumber\\\label{13}\\ B &=& \frac{11}{21}-\frac{11}{7}\nu +
 \left(\frac{1607}{378}-\frac{1681}{378}\nu+\frac{229}{378}\nu^2\right)
 \gamma \nonumber\\
 &+&\left(-\frac{357761}{19800}+\frac{428}{105}\ln\left(\frac{r}{r_0}\right)
 +\left[-\frac{75091}{5544}+\frac{44}{3}\zeta\right]\nu
 +\frac{35759}{924}\nu^2+\frac{457}{5544}\nu^3\right)
 \gamma^2\label{14}\;,
\end{eqnarray}
where the post-Newtonian ordering parameter (defined in harmonic
coordinates) reads

\begin{equation}\label{15}
\gamma=\frac{G M}{r c^2}\;.
\end{equation}
Notice the logarithms entering these formulas at the 3PN order. One
type involves a constant length scale $r_0$ coming from the general
wave-generation formalism of Refs. \cite{B95,B98tail,B98mult}, and
which corresponds to some ``infrared'' cut-off in the bound at
infinity of the integrals defining the multipole moments. The constant
$r_0$ should and will cancel out when considering the complete
multipole expansion of the field exterior to the source. The other
type of logarithm contains a different length scale $r_0'$, which
should rather be viewed as an ``ultra-violet'' cut-off associated with
the singular behaviour of the metric at the location of the
point-particles; $r'_0$ is related to the two constants $r'_1$ and
$r'_2$ --- one for each particles --- appearing in the 3PN equations
of motion of Refs. \cite{BF00,BFeom,BFreg,BFregM,ABF01,BI02CM} by

\begin{equation}\label{16}
\ln r'_0 = \frac{m_1}{M} \ln r'_1 + \frac{m_2}{M} \ln r'_2\;.
\end{equation}
As we know that $r'_1$ and $r'_2$ are gauge constants, i.e. they can
be eliminated by a gauge transformation at the 3PN order, $r'_0$ will
necessarily disappear from our physical, gauge-invariant, result at
the end. Besides the harmless constants $r_0$ and $r_0'$, there are
three point-mass regularization-constants in
Eqs. (\ref{13})-(\ref{14}): $\xi$, $\kappa$ and $\zeta$ (see Section
\ref{sec5}). However, we shall see that the energy flux for circular
orbits depends on one combination only of these constants:
$\theta=\xi+2\kappa+\zeta$.

Through 3.5PN order, the result is decomposed into an
``instantaneous'' part, i.e. generated solely by the multipole moments
of the source, and a tail part. What we call here the tail part is in
fact a complicated sum of ``tails'', ``tail squares'', and ``tails of
tails'', as computed in Ref. \cite{B98tail}. We find, respectively,

\begin{eqnarray}
 {\cal L}_{\rm inst} &=& {32c^5\over 5G}\nu^2 \gamma^5 \biggl\{ 1 +
\left(-\frac{2927}{336}-\frac{5}{4}\nu \right) \gamma +
\left(\frac{293383}{9072}+\frac{380}{9}\nu\right) \gamma^2 \nonumber
\\ &+&\left(\frac{53712289}{1108800} -\frac{1712}{105}\ln
\left(\frac{r}{r_0}\right)\right.\nonumber\\&+&\left.
\left[-\frac{332051}{720}+\frac{123}{64}\pi^2+\frac{110}{3}\ln
\left(\frac{r}{r_0'}\right) +44\lambda-\frac{88}{3}\theta\right]\nu
-\frac{383}{9}\nu^2\right) \gamma^3\biggr\}\;,\label{18}\\\nonumber\\
{\cal L}_{\rm tail} &=& {32c^5\over 5G}\nu^2 \gamma^5 \biggl\{ 4\pi
\gamma^{3/2}+\left(-\frac{25663}{672}-\frac{109}{8}\nu\right)\pi
\gamma^{5/2}\nonumber\\
&+&\left(-\frac{116761}{3675}+\frac{16}{3}\pi^2-\frac{1712}{105}C
-\frac{856}{105}\ln (16~\!\gamma)+\frac{1712}{105}\ln
\left(\frac{r}{r_0}\right)\right) \gamma^3\nonumber\\ &+&
\left(\frac{90205}{576} +\frac{3772673}{12096}\nu
+\frac{32147}{3024}\nu^2\right)\pi \gamma^{7/2} \biggr\}\;.\label{19}
\end{eqnarray}
The Newtonian result has been factorized out in front. Here
$C=0.577\cdots$ denotes the Euler constant. As we can see, the
constant $r_0$ drops out from the sum of Eqs. (\ref{18}) and
(\ref{19}). However, the gauge constant $r_0'$ does not seem to
disappear at this stage, but that is simply due to our use of the
post-Newtonian parameter $\gamma$ defined by (\ref{15}), and which
depends {\it via} the equation of motion on the choice of harmonic
coordinates. After substituting the frequency-related parameter $x$ in
place of $\gamma$ (with the help of the 3PN equations of motion), we
find that $r_0'$ does cancel as well --- this nicely illustrates the
consistency between our two computations, in harmonic-coordinates, of
the equation of motion on one hand and the multipole moments on the
other hand. Finally we obtain

\begin{eqnarray}\label{20}
 {\cal L} &=& {32c^5\over 5G}\nu^2 x^5 \biggl\{ 1 +
\left(-\frac{1247}{336}-\frac{35}{12}\nu \right) x + 4\pi
x^{3/2}\nonumber \\ &+&
\left(-\frac{44711}{9072}+\frac{9271}{504}\nu+\frac{65}{18}
\nu^2\right) x^2 +\left(-\frac{8191}{672}-\frac{535}{24}\nu\right)\pi
x^{5/2}\nonumber \\
&+&\left(\frac{6643739519}{69854400}+\frac{16}{3}\pi^2-\frac{1712}{105}C
-\frac{856}{105}\ln (16~\!x) \right.\nonumber\\
&+&\left.\left[-\frac{11497453}{272160}+\frac{41}{48}\pi^2
+\frac{176}{9}\lambda-\frac{88}{3}\theta\right]\nu-\frac{94403}{3024}\nu^2
-\frac{775}{324}\nu^3\right) x^3\nonumber\\ &+&
\left(-\frac{16285}{504}+\frac{176419}{1512}\nu
+\frac{19897}{378}\nu^2\right)\pi x^{7/2} \biggr\}\;.
\end{eqnarray}
The last test (but not the least) is that Eq. (\ref{20}) is in perfect
agreement, in the test-mass limit $\nu\to 0$, with the result of
linear black-hole perturbations obtained by Tagoshi and Sasaki
\cite{TSasa94}.

\section{Problem of regularization ambiguities}\label{sec5}

A model of structureless point masses is expected to be sufficient to
describe the inspiral phase of compact binaries. Thus we want to
compute the metric (and its gradient needed in the equations of
motion) at the 3PN order for a system of two point-like
particles. Applying general expressions for the metric valid in the
case of continuous (smooth) matter sources, we find that most of the
integrals become divergent at the location of the
particles. Consequently we must supplement the calculation by a
prescription for how to remove (i.e. to regularize) the infinite part
of these integrals.

The ``standard'' Hadamard regularization yields some ambiguous results
for the computation of certain integrals at the 3PN order, as
Jaranowski and Sch\"afer \cite{JaraS98,JaraS99} noticed in
their computation of the equations of motion within the
ADM-Hamiltonian formulation of general relativity. They showed that
there are {\it two} and {\it only two} types of ambiguous terms in the
3PN Hamiltonian, which were then parametrized by two unknown numerical
coefficients $\omega_{\rm static}$ and $\omega_{\rm kinetic}$.

Blanchet and Faye \cite{BFreg,BFregM}, motivated by the previous
result, introduced their ``improved'' Hadamard regularization, based
on a theory of pseudo-functions and generalized distributional
derivatives. This new regularization is mathematically well-defined
and free of ambiguities; in particular it yields unique results for
the computation of any of the integrals occuring in the 3PN equations
of motion. Unfortunately, this regularization turned out to be in a
sense incomplete, because it was found \cite{BF00,BFeom} that the 3PN
equations of motion involve {\it one} and {\it only one} unknown
numerical constant, called $\lambda$, which cannot be determined
within the method.  The comparison of this result with the work of
Jaranowski and Sch\"afer \cite{JaraS98,JaraS99}, on the basis
of the computation of the invariant energy of binaries moving on
circular orbits, showed \cite{BF00} that

\begin{eqnarray}
\omega_{\rm kinetic}&=&\frac{41}{24}\;,\label{21}\\
\omega_{\rm
static}&=&-\frac{11}{3}\lambda-\frac{1987}{840}\;.\label{22}
\end{eqnarray}
Therefore, the ambiguity $\omega_{\rm kinetic}$ is fixed, while
$\lambda$ is equivalent to the other ambiguity $\omega_{\rm
static}$. The value of $\omega_{\rm kinetic}$ given by Eq. (\ref{21})
was recovered by Damour, Jaranowski and Sch\"afer \cite{DJS00}, who
proved that this value is the unique one for which the global
Poincar\'e invariance of their formalism is verified. By contrast, the
harmonic-coordinate conditions preserve the Poincar\'e invariance, and
therefore the associated equations of motion should be
Lorentz-invariant, as was indeed found to be the case by Blanchet and
Faye \cite{BF00,BFeom}, thanks in particular to their use of a
Lorentz-invariant regularization \cite{BFregM} (hence their
determination of $\omega_{\rm kinetic}$).

The appearance of one and only one physical undeterminacy
$\lambda\Leftrightarrow\omega_{\rm static}$ in the equations of motion
constitutes a quite striking fact, specifically related to the use of
an Hadamard-type regularization. Mathematically speaking, the presence
of $\lambda$ is (probably) related to the fact that it is impossible
to construct a distributional derivative operator satisfying the
Leibniz rule for the derivation of the product. The Einstein field
equations can be written into many different forms, by operating some
terms by parts with the help of the Leibniz rule. All these forms are
equivalent in the case of regular sources, but they become
inequivalent for point particles if the derivative operator violates
the Leibniz rule.

In Ref. \cite{DJSisco} it has been argued that the numerical value of
the parameter $\omega_{\rm static}$ could be $\simeq -9$, because for
such a value some different resummation techniques, {\it viz} Pad\'e
approximants \cite{DIS98} and effective-one-body (EOB) method
\cite{BD01}, when they are implemented at the 3PN order, give
approximately the same numerical result for the location of the last
stable circular orbit. Even more, it was suggested \cite{DJSisco} that
$\omega_{\rm static}$ might be precisely equal to $\omega_{\rm
static}^*$, with

\begin{equation}\label{23}
\omega_{\rm static}^*=-\frac{47}{3}+\frac{41}{64}\pi^2\;.
\end{equation}
(We have $\omega_{\rm static}^*=-9.34\cdots$.) However, the value of
$\omega_{\rm static}$ in general relativity has been computed by
Damour, Jaranowski and Sch\"afer \cite{DJSdim} by means of a
dimensional regularization, instead of an Hadamard-type one, within
the ADM-Hamiltonian formalism. Their result is

\begin{equation}\label{24}
\omega_{\rm
static}=0~~\Longleftrightarrow~~\lambda=-\frac{1987}{3080}\;.
\end{equation}
As Damour {\it et al} \cite{DJSdim} argue, clearing up the ambiguity
is made possible by the fact that the dimensional regularization,
contrary to the Hadamard regularization, respects all the basic
properties of the algebraic and differential calculus of ordinary
functions\footnote{Note that the result (\ref{24}) is very different
from $\omega_{\rm static}^*$ given by Eq. (\ref{23}): this suggests
that the resummation techniques (Pad\'e approximants and EOB method),
though they are designed to ``accelerate'' the convergence of the
post-Newtonian series, do {\it not} in fact converge toward the same
exact solution (or, at least, not as fast as expected). See Section
\ref{sec7} for discussion on this point.}.

Let us comment that the use of a self-field regularization in this
problem, it be dimensional or based on the Hadamard partie finie,
signals a somewhat unsatisfactory situation on the physical point of
view, because we would like to perform, ideally, a complete
calculation valid for extended bodies, taking into account the details
of the internal structure of the bodies (energy density, pressure,
internal velocities, etc.). By considering the limit where the radii
of the objects tend to zero, one should recover the same result as
obtained by means of the point-mass regularization, and determine the
value of the regularization parameter $\lambda$. Because of
considerable difficulties arising at the 3PN order this program has
not yet been achieved.

Concerning the 3PN radiation field of two point masses --- the second
half of the problem, besides the 3PN equations of motion ---,
Blanchet, Iyer and Joguet \cite{BIJ02} used the (standard) Hadamard
regularization and found necessary to introduce three additional
regularization constants $\xi$, $\kappa$ and $\zeta$. However the
total gravitational-wave flux, in the case of circular orbits, depends
only on the linear combination 

\begin{equation}\label{25}
\theta = \xi+2\kappa+\zeta\;.
\end{equation}
Furthermore, this $\theta$ comes in at the same level as $\lambda$, so
there is in fact only one unknown constant in the flux [see
Eqs. (\ref{18})-(\ref{20})]. Notice that the improved version of the
Hadamard regularization proposed in Refs. \cite{BFreg,BFregM} should
be able, in principle, to fix the value of the constant $\zeta$.

\section{Accuracy of the post-Newtonian approximation}\label{sec6}

In this section we discuss (some aspects of) the accuracy of the
post-Newtonian approximation as regards the determination of the
binary's innermost circular orbit (ICO). The ICO will be defined by
the minimum, when it exists, of the energy function $E(x)$ given by
Eq. (\ref{10}). Notice that we do not define the ICO as an ISCO,
i.e. as a point of dynamical general-relativistic unstability. See
Ref. \cite{BI02CM} for a discussion of the dynamical stability of
circular binary orbits at the 3PN order.

Let us first confront the prediction of the standard (Taylor-based)
post-Newtonian approximation at the 3PN order for the ICO --- as given
by the minimum of Eq. (\ref{10}) --- with a recent numerical
calculation by Gourgoulhon, Grandcl\'ement and Bonazzola
\cite{GGB1,GGB2}. These authors obtained numerically the energy
$E(\omega)$ of binary black holes along evolutionary sequences of
equilibrium configurations under the assumptions of conformal flatness
for the spatial metric and of exactly circular orbits. The latter
restriction is implemented by requiring the existence of an
``helical'' Killing vector, time-like inside the light cylinder
associated with the circular motion and space-like outside. The
numerical calculation \cite{GGB1,GGB2} has been performed in the case
of {\it corotating} black holes, which are spinning with the orbital
angular velocity $\omega$. For the comparison we must therefore
include within the post-Newtonian formalism yielding Eq. (\ref{10})
the effects of spins appropriate to two Kerr black holes rotating at
the orbital rate $\omega$.

\begin{figure}[h]
\centerline{\epsfxsize=10cm \epsfbox{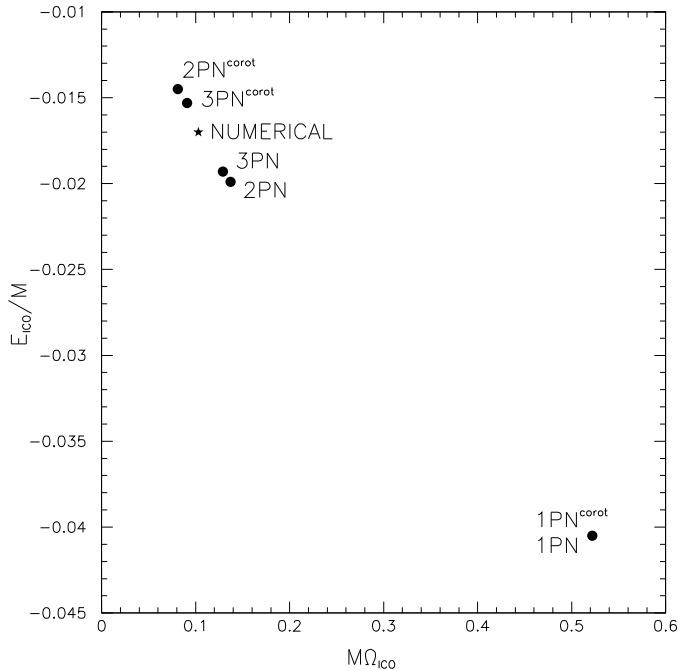}}
\vspace{0.5cm}
\caption{The center-of-mass energy $E_{\rm ICO}$ versus $\omega_{\rm
ICO}$ in the equal-mass case ($\nu=\frac{1}{4}$). The asterisk marks
the result calculated by numerical relativity \cite{GGB1,GGB2}. The points
indicated by 1PN, 2PN and 3PN correspond to irrotational binaries,
while the points denoted by 1PN$^{\rm corot}$, 2PN$^{\rm corot}$ and
3PN$^{\rm corot}$ describe corotational binaries. Both 3PN are
3PN$^{\rm corot}$ are shown for $\omega_{\rm static}=0$.}
\label{fig}
\end{figure}

The Figure \ref{fig} (issued from Ref. \cite{B02ico}) presents the
post-Newtonian results for $E_{\rm ICO}$ in the case of irrotational
and corotational binaries. The points indicated by 1PN, 2PN and 3PN
are defined from the obvious truncation of Eq. (\ref{10}). The points
1PN$^{\rm corot}$, 2PN$^{\rm corot}$ and 3PN$^{\rm corot}$ take into
account the spin effects of corotational binaries and are computed in
Ref. \cite{B02ico}. Notice that the irrotational and corotational
configurations differ only from the 2PN order. As we can see the 3PN
points, and even the 2PN ones, are rather close to the numerical value
(marked by an asterisk in Figure \ref{fig}). As expected, the best
agreement is for the 3PN approximation and in the case of corotation:
i.e. the point 3PN$^{\rm corot}$. So our first conclusion is that the
location of the ICO as computed by numerical relativity, under the
helical-symmetry approximation \cite{GGB1,GGB2}, is in good agreement
with post-Newtonian predictions. This constitutes an appreciable
improvement of the previous situation, because we recall that the
earlier estimates of the ICO in post-Newtonian theory \cite{KWW} and
numerical relativity \cite{Pfeiffer,Baumgarte} strongly disagree with
each other, and do not match with the present 3PN results (see
Ref. \cite{GGB2} for further discussion).

Our second conclusion comes from the fact that the 2PN and 3PN values
in Figure \ref{fig} --- either for irrotational or corotational
binaries --- are so close to each other. Indeed it seems that the 3PN
points are useful only to {\it confirm} the result given by the 2PN
ones. This is a quite satisfying state of affairs because it indicates
that the post-Newtonian approximation converges very well (in the
sense of Cauchy's criterion). Figure \ref{fig} shows that the
post-Newtonian approximation is well qualified to accurately locate
the ICO (of course for this purpose one must go to the 3PN order ---
the 1PN approximation is clearly not accurate enough).

Let us elaborate more on this point. First we make a few
order-of-magnitude estimates. At the location of the ICO we find (see
Figure \ref{fig}) that the frequency-related parameter $x$ defined by
Eq. (\ref{9}) is approximately of the order of 20\%. Therefore, we
might {\it a priori} expect that the contribution of the 1PN
approximation to the energy at the point of the ICO should be of the
order of 20\%. For the present discussion we take the pessimistic view
that the order of magnitude of an approximation represents also the
order of magnitude of the higher-order terms that are neglected. We
see that the 1PN approximation should yield a rather poor estimate of
the ``exact'' result, but this is quite normal at this very
relativistic point where the orbital velocity is $\frac{v}{c}\sim
\sqrt{x}\sim 50\%$. By the same argument we infer that the 2PN
approximation should do much better, with fractional errors of the
order of $x^2\sim 5\%$, while the 3PN approximation will be even
better, with the precision $x^3\sim 1\%$.

The simple order-of-magnitude estimate suggests therefore that the 3PN
order should be close to the ``exact'' solution for the ICO to within
1\% of fractional accuracy. We think that this is very good, and we
should even remember that this estimate is pessimistic, because we
could reasonably expect that the neglected higher-order
approximations, 4PN and so on, are in fact much smaller numerically
(e.g. of the order of $x^4\sim 0.2\%$). But let us keep for the
present discussion the 1\% guess for the accuracy of the 3PN
approximation.

Now the previous estimates make sense only if the numerical values of
the post-Newtonian coefficients in Eq. (\ref{10}) are roughly of the
order of one. If this is not the case, and if the coefficients
increase dangerously with the post-Newtonian order $n$, one sees that
the post-Newtonian approximation might in fact be very bad. So let us
look at the values of the coefficients of the 1PN, 2PN and 3PN
approximations in Eq. (\ref{10}), say

\begin{eqnarray}
a_1(\nu) &=& -\frac{3}{4}-\frac{\nu}{12}\;,\label{26}\\ a_2(\nu) &=&
-\frac{27}{8}+\frac{19}{8}\nu -\frac{\nu^2}{24}\;,\label{27}\\
a_3(\nu) &=&
-\frac{675}{64}+\left[\frac{209323}{4032}-\frac{205}{96}\pi^2
-\frac{110}{9}\lambda\right]\nu-\frac{155}{96}\nu^2
-\frac{35}{5184}\nu^3\;.\label{28}
\end{eqnarray} 
We present in Table \ref{tab1} the values of these coefficients in the
test-mass limit $\nu=0$, and in the equal-mass case $\nu=\frac{1}{4}$
when the ambiguity parameter takes the ``uncorrect'' value
$\omega^*_{\rm static}$ [see Eq. (\ref{23})] and the correct one
$\omega_{\rm static}=0$ predicted by general relativity \cite{DJSdim}.

\begin{table}[h]
\begin{center}
\footnotesize
\begin{tabular}{lccccc}
&Newtonian&$a_1(\nu)$&$a_2(\nu)$&$a_3(\nu)$\\[1mm]\hline\hline $\nu=0$
&1&-0.75&-3.37&-10.55\\[0.5mm]\hline $\nu=\frac{1}{4}$ $\quad\omega^*_{\rm
static}\simeq -9.34$&1&-0.77&-2.78&-8.75\\[0.5mm]\hline $\nu=\frac{1}{4}$
$\quad\omega_{\rm static}=0$&1&-0.77&-2.78&-0.97\\
\end{tabular}
\end{center}
\caption{Sequence of coefficients of the post-Newtonian series
composing the energy function (\ref{26})-(\ref{28}).\label{tab1}}
\end{table}

Our first comment is that when $\nu=0$ there is an {\it increase} of
the coefficients by roughly a factor 3 at each step. This behaviour is
fairly easy to understand. It comes from the existence in the
Schwarzshild metric of the famous light-ring orbit: a geodesics of
photon which is a circular orbit located at $R=3M$ in Schwarzshild
coordinates. As a result, the energy of the test particle in the
Schwarzshild metric exhibits a singularity at $x_{\rm
light-ring}=\frac{1}{3}$,

\begin{equation}\label{29}
E^{\rm test}(x) = \mu c^2\left(\frac{1-2x}{\sqrt{1-3x}}-1\right)\;.
\end{equation}
The consequence is that the radius of convergence of the
post-Newtonian series is $\frac{1}{3}$, so the post-Newtonian
coefficients increase by a factor $\sim 3$. So the post-Newtonian
series is not very accurate in the case $\nu=0$. This fact has
motivated several statements in the literature (see
e.g. \cite{3mn,CFPS93,P95,DIS98}), according to which the
post-Newtonian approximation would be ``poorly convergent'', or that
there should be a ``fundamental breakdown'' of its validity in the
regime of the ICO. This is indeed true in the {\it test-mass} limit
($\nu\to 0$), where the post-Newtonian series converges
slowly\footnote{Let us remark that this negative conclusion does not
matter: indeed we shall never use the post-Newtonian approximation in
the case $\nu\to 0$ simply because we know the exact result which is
given by Eq. (\ref{10}). The exact result for the radiation field is
known as well, albeit numerically only \cite{TNaka94,P95}. Therefore
we should not worry too much about the poor convergence of the
post-Newtonian series in the test-mass limit. The post-Newtonian
method is useless and even one might say irrelevant when considering
the motion of a test particle around a Schwarzschild black hole.}.

On the other hand, what happens in the equal-mass case
($\nu=\frac{1}{4}$)? When $\nu=\frac{1}{4}$ and we have the value
$\omega^*_{\rm static}$ we notice that the coefficients increase
approximately like in the test-mass case $\nu=0$. This indicates that
the gravitational interaction in the case of $\omega^*_{\rm static}$
looks like that in a one-body problem. We shall say that when the
post-Newtonian coefficients rapidly increase with the order of
approximation, the interaction is ``Schwarzschild-like''.

Now when $\nu=\frac{1}{4}$ and the ambiguity parameter takes the
correct value $\omega_{\rm static}=0$, we see that the 3PN coefficient
$a_3(\frac{1}{4})$ is of the order of minus one instead of $\sim
-10$. We think that this strongly suggests, unless 3PN happens to be
quite accidental, that the post-Newtonian coefficients in general
relativity do not increase very much with $n$, and stay rather of the
order of one. This is interesting as it indicates that the actual
general-relativistic two-body interaction is {\it not}
Schwarzschild-like.

It is impossible of course to be very confident about the validity of
the previous statement because we know only the coefficients up to the
3PN order. Any tentative conclusion based on the 3PN order can be
``falsified'' when we obtain the next 4PN order. Nevertheless, we feel
that the mere fact that $a_3(\frac{1}{4})=-0.97$ in Table \ref{tab1} is
sufficient to motivate our (tentative) conclusion that the field of
two bodies is more complicated than the Schwarzschild space-time. This
conclusion is in accordance with the present author's respectfulness
of the complexity of the Einstein field equations.

The nice consequence is that because the post-Newtonian coefficients
when $\nu=\frac{1}{4}$ stay of the order of one, the {\it standard}
post-Newtonian approach, based on the standard Taylor approximants, is
probably very accurate. The post-Newtonian series seems to ``converge
well'', with a ``convergence radius'' of the order of
one\footnote{Actually, the post-Newtonian series could be only
asymptotic (hence divergent), but nevertheless it should give good
results provided that the series is truncated near some optimal order
of approximation. In this discussion we assume that the 3PN order is
not too far from that optimum.}. A convincing support of this view is
provided by Figure \ref{fig} itself. The order-of-magnitude estimates
we did at the beginning of this Section are probably correct. In
particular the 3PN order should be close to the ``exact'' solution
even in the regime of the ICO.

It is also interesting to look at the numerical values of the
post-Newtonian coefficients in the total flux ${\cal L}$ obtained in
Eq. (\ref{20}). For this case we do not have (like for the ICO) a
clear point at which we can compare with a result of numerical
relativity. Furthermore things concerning the convergence of the
post-Newtonian series are less clear than with the energy function. A
possible reason for that could be that the flux, in contrast to the
energy, includes the contribution of tails at the 1.5PN, 2.5PN and
3.5PN orders, and even of tails-of-tails at the 3PN order.

\begin{table}[h]
\begin{center}
\footnotesize
\begin{tabular}{lcccccccc}
&Newtonian&$b_1(\nu)$&$b_{3/2}(\nu)$&$b_2(\nu)$&$b_{5/2}(\nu)$&
$b_3(\nu)$&$b_{7/2}(\nu)$\\[1mm]\hline\hline $\nu=0$
&1&-3.71&12.57&-4.93&-38.29&128.85&-101.51\\[0.5mm]\hline
$\nu=\frac{1}{4}$
$\quad\lambda=-\frac{1987}{3080}\quad\theta=0$&1&-4.44&12.57
&-0.10&-55.80&115.26&0.47\\
\end{tabular}
\end{center}
\caption{Post-Newtonian coefficients in the flux function
(\ref{20}). The coefficient $b_3(\nu)$ contains a log-term that we
have evaluated at $x_{\rm ICO}\sim 0.2$. [The value $\theta=0$ is
taken to be indicative.] \label{tab2}}
\end{table}

\section{On Schwarzschild-like templates for binary inspiral}\label{sec7}

Let us finally comment about a possible implication of our conclusion
as regards the validity of the so-called post-Newtonian resummation
techniques, i.e. Pad\'e approximants \cite{DIS98,DJSisco}, which aim
at ``accelerating'' the convergence of the post-Newtonian series in
the pre-coalescence stage, and effective-one-body (EOB) methods
\cite{BD01,DJSisco}, which attempt at describing the late stage of the
coalescence of two black holes. These techniques are based on the idea
that the gravitational two-body interaction is a ``deformation'' ---
with $\nu\leq\frac{1}{4}$ being the deformation parameter --- of the
Schwarzschild space-time. 

The Pad\'e approximants are valuable tools for giving accurate
representations of functions having some singularities. In the problem
at hands they would be justified if the ``exact'' expression of the
energy [whose 3PN expansion is given by Eqs. (\ref{26})-(\ref{28})]
would admit some Schwarzschild-like features, and in particular a
light-ring singularity. In the Schwarzschild case the Pad\'e series
converges rapidly toward the solution \cite{DIS98}: the Pad\'e
constructed only from the 2PN approximation of the energy --- keeping
only $a_1(0)$ and $a_2(0)$ in Eqs. (\ref{2}) --- already coincide with
the exact result given by Eq. (\ref{8}). On the other hand, the EOB
method maps the post-Newtonian two-body dynamics (at the 2PN or 3PN
order) on the geodesic motion on some effective metric which happens
to be a $\nu$-deformation of the Schwarzschild space-time. In the EOB
method the effective metric looks like Schwarzschild {\it by
definition}, and we might expect the two-body interaction to own some
Schwarzschild-like features.

Our comment is that the validity of these resummation techniques (and
the templates for binary inspiral built on them) is questionable,
because as we have seen in Section \ref{sec6} the value $\omega_{\rm
static}=0$ suggests that most probably the two-body interaction is not
Schwarzschild-like. In particular, there does not seem to exist
something like a light-ring orbit which would be a deformation of the
Schwarzschild one.

To test the previous idea, let us come back to the Schwarzschild
limit, where we observed that the radius of convergence of the
post-Newtonian series is given by the light-ring singularity at the
value $\frac{1}{3}$. Now the radius of convergence is given by
d'Alembert's criterion as the limit when $n\to +\infty$ of the ratio
of successive coefficients, i.e. $a_{n-1}/a_n$. We might therefore
{\it compute} the light-ring orbit by investigating the limit

\begin{equation}\label{30}
x^{\rm test}_{\rm light-ring} = \lim_{n\to
+\infty}\,\frac{a_{n-1}^{\rm test}}{a_n^{\rm test}} \,=\,
\frac{1}{3}\;.
\end{equation}
To test for the possible existence of a light-ring singularity in
the case of comparable masses, we consider the ratio between the two
highest known post-Newtonian coefficients, that are $a_2(\nu)$ and
$a_3(\nu)$. This ratio will give an {\it estimate} of the
``light-ring'' singularity for non-infinitesimal mass ratios,

\begin{equation}\label{31}
x_{\rm light-ring}(\nu)\sim\frac{a_2(\nu)}{a_3(\nu)} \;.
\end{equation}
Using the values given in Table \ref{tab1} we obtain for equal masses
in the case of the ``wrong'' ambiguity parameter $\omega^*_{\rm
static}\simeq -9.34$,

\begin{equation}\label{32}
x_{\rm light-ring}(\hbox{$\frac{1}{4}$},{\omega^*_{\rm static}})\sim
0.32\;.
\end{equation}
As we see there seems to be in this case a (pseudo-)light-ring orbit
which is a small deformation of the Schwarzschild light-ring orbit
given by Eq. (\ref{30}). In this Schwarzschild-like situation, we
expect that Pad\'e approximants and EOB-type methods to be
appropriate. 

{\it But}, in the case of general relativity ($\omega_{\rm
static}=0$), we obtain a drastically different result:

\begin{equation}\label{33}
x_{\rm light-ring}(\hbox{$\frac{1}{4}$},G.R.) \sim 2.86\;.
\end{equation}
If we believe the correctness of this estimate we must conclude that
there is in fact no notion of a light-ring orbit in the real two-body
case. Or, one might say (pictorially speaking) that the light-ring
orbit gets hidden inside the horizon of the final black-hole formed by
coalescence. If we consider the ratio between the 1PN and 2PN
coefficients (instead of between 2PN and 3PN), we get the value $\sim
0.28$ instead of Eq. (\ref{33}). So at the 2PN order the field seems
to admit a light ring, while at the 3PN order it does not. This
reinforces our idea that it is meaningless (with our present 3PN-based
knowledge, and untill fuller information is available), to assume the
existence of a light-ring singularity in the equal-mass case. Our
expectation, therefore, is that the conditions under which the Pad\'e
and EOB methods should be legitimate might in fact not be fulfilled.

This doubt is confirmed by the finding of Ref. \cite{DJSisco} (already
alluded to above) that in the case of the ``wrong'' ambiguity
parameter $\omega^*_{\rm static}\simeq -9.34$ the Pad\'e approximants
and the EOB method at the 3PN order give the same result for the
ICO. From the previous discussion we see that this agreement is to be
expected because a deformed (pseudo-)light-ring singularity seems to
exist with $\omega^*_{\rm static}$. By contrast, in the case of
general relativity, where $\omega_{\rm static}=0$, the Pad\'e and EOB
methods give quite different results ({\it cf} the figure 2 in
Ref. \cite{DJSisco}). Another confirmation comes from the light-ring
singularity which is determined from the Pad\'e approximants at the
2PN order (see Eq. (3.22) in Ref. \cite{DIS98}) as $x^{\rm P-2PN}_{\rm
light-ring}(\frac{1}{4})\sim 0.44$. This value is rather close to
Eq. (\ref{32}) but strongly disagrees with Eq. (\ref{33}). Our
explanation is that the Pad\'e series might converge toward a theory
having $\omega_{\rm static}=\omega^*_{\rm static}$ and therefore which
is different from general relativity.

\end{document}